\documentclass[twocolumn,aps,pra,showpacs,amsmath,superscriptaddress]{revtex4}
\usepackage{amssymb}
\usepackage{mathrsfs}
\usepackage{graphicx}
\usepackage{float}

\begin{document}

\title{Bose-Fermi mixture in one-dimensional optical lattices with hard-core interactions}
\author{Xiaoming Cai}
\affiliation{Beijing National Laboratory for Condensed Matter
Physics, Institute of Physics, Chinese Academy of Sciences, Beijing
100190, China}

\author{Liming Guan}
\affiliation{Beijing National Laboratory for Condensed Matter
Physics, Institute of Physics, Chinese Academy of Sciences, Beijing
100190, China}

\author{Shu Chen}
\email{schen@aphy.iphy.ac.cn} \affiliation{Beijing National
Laboratory for Condensed Matter Physics, Institute of Physics,
Chinese Academy of Sciences, Beijing 100190, China}

\author{Yajiang Hao}
\affiliation{Department of Physics, University of Science and
Technology Beijing, Beijing 100083, China}

\author{Yupeng Wang}
\affiliation{Beijing National Laboratory for Condensed Matter
Physics, Institute of Physics, Chinese Academy of Sciences, Beijing
100190, China}
\date{ \today}

\begin{abstract}
We study a mixture of $N_{b}$ bosons with point hard-core
boson-boson interactions and $N_{f}$ noninteracting spinless
fermions with point hard-core boson-fermion interactions in 1D
optical lattice with external  harmonic confine potential. Using an
extended Jordan-Winger transformation (JWT) which maps the hard-core
Bose-Fermi mixture into two component noninteracting spinless
fermions with hard-core interactions between them, we get the ground
states of the system. Then we determine in details the one particle
density matrix, density profile, momentum distribution, the natural
orbitals and their occupations based on the constructed ground state
wavefunctions. We also discuss the ground state properties of the
system with large but finite interactions which lead to the lift of
ground degeneracy. Our results show that, although the total density
profile is almost not affected, the distributions for bosons and
fermions strongly depend on the relative strengthes of boson-boson
interactions and boson-fermion interactions.

\end{abstract}

 \maketitle

\section{Introduction}
Recently, strongly interacting gases of bosons in one dimension have
been experimentally realized \cite{Stoeferle,Paredes,Toshiya} by
loading a Bose-Einstein condensate into a deep two dimensional
optical lattice to create arrays of one-dimensional (1D) atomic
systems. The achievement of the Tonks-Girardeau (TG) regime
\cite{Girardeau} in an optical lattice has stimulated extensive
theoretical interest in the study of many-body physics of 1D quantum
gases \cite{Olshanii,Petrov,Dunjko,Chen}. With the increase in the
interaction strength, the 1D Bose gas evolves from a Bose-Einstein
condensate to ``fermionized" TG
\cite{LL,Hao06,Hao07,Zoellner,Deuretzbacher,Landau,Alon}. The
microscopic mechanisms of the evolution has been recently studied by
both analytical methods \cite{Hao06,Hao07} and various numerical
methods \cite{Alon,Zoellner,Deuretzbacher,Landau}. In the limit of
the infinitely repulsive interaction, the many-body state of a TG
gas has been shown to correspond to the states of a noninteracting
Fermi gas via a Bose-Fermi mapping \cite{Girardeau}. For the lattice
model of TG gas, the Hamiltonian in a periodic lattice can be mapped
onto the 1D XY model of Lieb, Schulz, and Mattis, which has been
extensively studied in the literature \cite{Lieb}. With an
additional confining potential, the lattice TG gas has been studied
by means of an exact numerical approach by Rigol and Muramatsu
\cite{Rigol,Rigol1,Rigol2}. Very recently, Girardeau's Bose-Fermi
mapping method has also been generalized to deal with mixtures of
multi-component quantum gases
\cite{Girardeau07,Deuretzbacher08,GuanLM,Fang}.

On the other hand, mixtures of bosonic and fermionic atoms have been
studied extensively as they initially provided a convenient way to
achieve degenerate fermionic gas by means of sympathetic cooling
\cite{Truscott,Modugno}. Due to their rich phase diagram, the
Bose-Fermi mixtures have attracted many theoretical studies
\cite{Das,Cazalilla,Lewenstein,Mathey,Albus,Demler,GuanXW,Hebert,
Senguptal,Troyer,Roth,Zujev,Kuklov,Frahm,Lai,Yin,Cramer}. Among
those works, particular attention has been paid to the 1D model of
mixed bosons and polarized fermions
\cite{Das,Cazalilla,Lewenstein,Mathey,Albus,Demler,GuanXW,Hebert,Yin},
in which there are only s-wave scattering for boson-boson and
boson-fermion interactions present. While most of these
investigations relied on the mean-field approximations or the
Luttinger liquid theory, there is rarely exact result except for the
homogenous model with equal boson-boson and boson-fermion
interactions, which is exactly solved by the Bethe-ansatz method
\cite{Demler,GuanXW,Yin,Frahm,Lai}. However, for a lattice system,
there is no exact result even for the lattice correspondence of the
integrable continuum Bose-Fermi system \cite{Cazalilla04}. In this
work we mainly study the lattice Bose-Fermi gas in the hard-core
limit where both the boson-boson and boson-fermion interactions are
infinitely strong. In this limit, we can apply an extended
Jordan-Wigner transformation and an exact numerical approach, which
can be viewed as an extension of the method by Rigol and Muramatsu
\cite{Rigol}, to treat the hard-core Bose-Fermi mixture exactly. We
focus on the ground state properties and analyze the behaviors of
the one-particle correlations, the momentum distribution function,
the natural orbitals and their occupations. The properties of the
system with large but finite interactions are also discussed. Since
the interaction can be tuned in principle within a large range of
regime by exploiting Fesbach resonance, our results might be
experimentally relevant.

The content of the paper is as follows. In the next section we
describe the model system first and then the approach used to
calculate the ground state one particle density matrix exactly. In
Sec. III, we discuss the properties of hard-core Bose-Fermi mixture
confined in harmonic traps. In the Sec. IV, we discuss the
properties of the system with large but finite interactions. The
paper is concluded in Sec. V.

\section{System and method}
We consider a mixture of $N_{b}$ bosons with point hard-core
boson-boson interactions and $N_{f}$ noninteracting spinless
fermions with point hard-core boson-fermion interactions and
assume that the boson and fermion particles have the same masses
$m_{b}=m_{f}$, which could be realized by choosing an isotope of a
given alkali element. Let $ X_{b}=(x_{1b},...,x_{N_{b}b})$ and
$X_{f}=(x_{1f},...,x_{N_{f}f})$ indicate the boson and fermion
coordinates respectively. For the system trapped in the potentials
including the optical lattice and an additional harmonic trap, the
Hamiltonian of the system is
\begin{equation}
H=H_{B}+H_{F}+H_{BB}+H_{BF}
\end{equation} with
\begin{eqnarray}
H_{B}&=&\sum^{N_{b}}_{j=1}\left[\frac{-\hbar^{2}}{2m}%
\frac{\partial ^{2}}{\partial x_{jb}^{2}}+v(x_{jb})\right] \notag\\
H_{F}&=&\sum^{N_{f}}_{j=1}\left[ \frac{-\hbar ^{2}}{2m}%
\frac{\partial ^{2}}{\partial x_{jf}^{2}}+v(x_{jf})\right] \notag\\
H_{BB}&=&\frac{g_{BB}}{2}\sum^{N_{b}}_{i,j}\delta
(x_{ib}-x_{jb}) \notag\\
H_{BF} &=&g_{BF}\sum^{N_b,N_{f}}_{i,j}\delta (x_{ib}-x_{jf})
\end{eqnarray}
where $g_{_{BB}}, \, g_{_{BF}}\rightarrow \infty $ under the
hard-core condition which means that the many-body wave function
$\Psi (X_{b},X_{f})$ of the system vanishes at all boson-boson (BB)
and boson-fermion (BF) collision points. Explicitly, the potentials
take the form
\[
v_{b,f}(x)=V_{b,f}^0 \sin^2(\pi x/ a)+\frac{1}{2} m_{b,f}
\omega_{b,f}^2 x^2
\]
where $a$ is the lattice spacing associated with wave vector $k_L=
\pi/a$ of the standing laser light. In this work, we consider only
the case with the trap acting on bosons and fermions being the
same, ie., $v_{b}(x)=v_{f}(x)=v(x)$ with $V_{b}^0=V_{f}^0$ and
$\omega_{b}=\omega_{f}$.

We use the Wannier function (only consider the lowest Bloch band) of
the optical lattice to expand $\Psi (X_{b},X_{f})$ and get  the
second quantized Hamiltonian of $H$, which is the standard Hubbard
model of Bose-Fermi mixture \cite{Albus} with the from of
\begin{eqnarray}
\label{eqn8}
H_{Hub}&=&-t\sum^{L-1}_{i=1}(b^\dagger_ib_{i+1}+f^\dagger_if_{i+1}+H.c.)\notag\\
&&+Va^{2}\sum^{L}_{i=1}i^{2}n^b_{i}+Va^{2}\sum^{L}_{i=1}i^{2}n^f_{i}\notag\\
&&+\frac{U_{bb}}{2}\sum^L_{i=1}n^b_i(n^b_i-1)+U_{bf}\sum^L
_{i=1}n^b_in^f_i \label{BFHubbard}
\end{eqnarray}
where $b_i^{\dagger}$ ($f_i^{\dagger}$) and $b_i$ ($f_i$) denote the
bosonic (fermionic) creation and annihilation operators at site i,
respectively, and they satisfy the standard (anti-) commutation
relations, i.e., $[b_{i},b_{j}^{\dagger}]=\delta _{ij}$,
$\{f_{i},f_{j}^{\dagger}\}=\delta _{ij}$, and $[b_{i},f_{j}]=0$. In
the hard-core limit, the Hamiltonian is simplified to
\begin{equation}
H=H_{b}+H_{f}
\end{equation}
with
\begin{eqnarray}
\label{citeeq1}
H_{b} &=&-t\sum^{L-1}_{i=1}(b_{i}^{\dagger}b_{i+1}+H.c.)+Va^{2}%
\sum^{L}_{i=1}i^{2}n^b_{i}\notag \\
H_{f} &=&-t\sum^{L-1}_{i=1}(f_{i}^{\dagger}f_{i+1}+H.c.)+Va^{2}%
\sum^{L}_{i=1}i^{2}n^f_{i}
\end{eqnarray}%
where additional on-site constraint
\[b^\dagger_ib_i+f^\dagger_if_i\leq1\]
is assigned to avoid double or higher occupancy \cite{note1}. Here
$t$ is the hopping parameter to be decided by the optical lattice;
$L$ is the number of the sites; $V(i)=Va^{2}i^{2}$ is the harmonic
confined potential with $a$ is the lattice space and $V$ is the
strength; $ n^b_{i}=b_{i}^{\dagger}b_{i}$ and
$n^f_{i}=f_{i}^{\dagger}f_{i}$ are the boson and fermion particle
number operators respectively. Given the local Hilbert space at
$i$-th site composed of a set of
$\{|0\rangle,b^\dagger_i|0\rangle,f^\dagger_i|0\rangle\}$ under the
single occupied on-site constraint, the on-site constraint can be
written in the follow forms
\begin{eqnarray}
\{b_i,b^\dagger_i\}&=&1-f^\dagger_if_i\notag\\
\{f_i,f^\dagger_i\}&=&1-b^\dagger_ib_i,
\end{eqnarray}
and the following equations are also valid
\begin{eqnarray}
b_{i}^{\dagger2} &=&b_{i}^{2}=f_{i}^{\dagger2} =f_{i}^{2}=0,\notag\\
b_{i}^{\dagger}f_{i}^{\dagger}
&=&f_{i}b_{i}=f_ib^\dagger_i=b_if^\dagger_i=0.
\end{eqnarray}

In order to get the ground state properties of the system, we extend
the general Jordan-Wigner transformation (JWT)\cite{Jordan} and get
the following transformations:
\begin{eqnarray}
\label{JWT} f_{j}^{\dagger}&=& \prod^{j-1}_{\beta =1}e^{-i\pi
c_{\beta\uparrow }^{\dagger}c_{\beta\uparrow
}}c_{j\uparrow}^{\dagger}, \notag\\
f_{j} &=& c_{j\uparrow}\prod^{j-1}_{\beta =1}
e^{+i\pi c_{\beta\uparrow}^{\dagger}c_{\beta\uparrow}} , \\
b_{j}^{\dagger} &=& \prod^{j-1}_{\beta =1}e^{-i\pi(
c_{\beta\downarrow }^{\dagger}c_{\beta\downarrow
}+c_{\beta\uparrow}^{\dagger}c_{\beta\uparrow})}c_{j\downarrow}^{\dagger}, \notag\\
b_{j} &=& c_{j\downarrow}\prod^{j-1}_{\beta =1} e^{+i\pi
(c_{\beta\downarrow }^{\dagger}c_{\beta\downarrow
}+c_{\beta\uparrow}^{\dagger}c_{\beta\uparrow})},\notag
\end{eqnarray}%
which map the Hamiltonian of the bosons into noninteracting spinless
fermions Hamiltonian. Using the JWT we can change the Hamiltonian of
the system into \[ H_{1}=H_{c\uparrow}+H_{c\downarrow} \] with
\begin{eqnarray}
H_{c\sigma}
&=&-t\sum^{L-1}_{i=1}(c_{i\sigma}^{\dagger}c_{i+1,\sigma}+H.c.)+Va^{2}
\sum^{L}_{i=1}i^{2}n^c_{i\sigma},
\end{eqnarray}%
where $\sigma=\uparrow,\downarrow$, and%
\[\{c_{i\uparrow},c^\dagger_{j\uparrow}\}=\{c_{i\downarrow},c^\dagger_{j\downarrow}\}=\{c_{i\downarrow},c_{j\uparrow}\}=0\]
for $i\neq j$, else
\begin{eqnarray}
\label{eqn1}
\{c_{i\downarrow},c^\dagger_{i\downarrow}\}=1-c^\dagger_{i\uparrow}c_{i\uparrow}&,&\{c_{i\uparrow},c^\dagger_{i\uparrow}\}=1-c^\dagger_{i\downarrow}c_{i\downarrow},\notag\\
c^{\dagger2}_{i\sigma}&=&c^2_{i\sigma}=0,\notag\\
c_{i\uparrow}c_{i\downarrow}=c^\dagger_{i\uparrow}c^\dagger_{i\downarrow}&=&c_{i\uparrow}c^\dagger_{i\downarrow}=c_{i\downarrow}c^\dagger_{i\uparrow}=0
\end{eqnarray}
for the  on-site constraints. Here
$n^{c}_{i\sigma}=c_{i\sigma}^\dagger c_{i\sigma}$ is the
$\sigma$-kind (we note the spinless fermions from the boson by
$\downarrow$-kind and the original fermions by $\uparrow$-kind)
fermion number operator, and $N_{\downarrow(\uparrow)}=N_{b(f)}$.
The Hamiltonian $H_{1}$ describes a mixture of two component
fermions with point hard-core interactions between two kinds. Notice
that the operators anticommute between two kinds.

Next we construct the ground state of the Hamiltonian $H_1$ under
the constraints Eq.(\ref{eqn1}) with the method proposed by Batista
{\it et al}  \cite{Batista}. We consider a set of parent states,
labeled by the string configuration $\vec{\mathbf{\sigma}}$, with
$N=N_\uparrow+N_\downarrow$ particles and $L-N$ holes,
$|\Phi_0(\vec{\sigma})\rangle$, and the form is:
\begin{equation}
|\Phi_0(\vec{\sigma})\rangle=|\underbrace{\sigma_1\sigma_2\sigma_3\cdot\cdot\cdot\sigma_{N}}_N\underbrace{\circ\circ\circ\cdot\cdot\cdot}_{L-N}\rangle,
\end{equation}
where $\sigma_i$ indicates the kind ($\uparrow$ or $\downarrow$) of
the fermion particle at site $i$, $L$ is the number of sites. Notice
that the number of  the configuration $\vec{\sigma}$ is $C^{N_b}_N$.
Then we rewrite the Hamiltonian $H_1$ with $H_1=T+H_V$, and
\begin{eqnarray}
T&=&-t\sum_{i,\sigma}T_{i\sigma},\quad T_{i\sigma}=c^\dagger_{i\sigma}c_{i+1,\sigma}+H.c.,\notag\\
H_V&=&Va^2\sum_{i,\sigma}i^2n^c_{i\sigma}.
\end{eqnarray}
The states $|\Phi_0(\vec{\sigma})\rangle$ are eigenstates of $H_V$
and they are degenerate with different $\vec{\sigma}$.

By applying the hopping operator $T_{i\sigma}$ we can generate a
subspace $M(\vec{\sigma})$ from the parent state
$|\Phi_0(\vec{\sigma})\rangle$, and we denote
\[|\Phi_1(\vec{\sigma})\rangle=T_{N,\sigma}|\Phi_0(\vec{\sigma})\rangle\]
or, in general
\[|\Phi_r(\vec{\sigma})\rangle=T_{i\sigma}|\Phi_j(\vec{\sigma})\rangle.\]
Obviously the dimension of the subspace $M(\vec{\sigma})$ is
$C^N_L$, and there are $C^{N_b}_N$ subspaces. Moreover these
different subspaces for different $\vec{\sigma}$ are orthogonal.

Next we construct the ground state in the subspace
$M(\vec{\sigma})$. For a specific $\vec{\sigma}$, we can make the
following mapping:
\begin{equation}
\label{mapping}
|\underbrace{\sigma_1\sigma_2\sigma_3\cdot\cdot\cdot\sigma_{N}}_N\underbrace{\circ\circ\circ\cdot\cdot\cdot}_{L-N}\rangle\rightarrow
|\underbrace{\bullet\bullet\bullet\cdot\cdot\cdot\bullet}_N\underbrace{\circ\circ\circ\cdot\cdot\cdot}_{L-N}\rangle
\end{equation}
which maps the two component fermions$(c_{i\sigma})$ into a single
spinless fermion $(c_i)$. It is straightforward to show that in the
corresponding new basis the system Hamiltonian can be written as
\begin{equation}
H_{spinless}=-t\sum^{L-1}_{i=1}(c^\dagger_i
c_{i+1}+H.c.)+Va^2\sum^L_{i=1}i^2n_i
\end{equation}

The ground state properties of the fermionic system $H_{spinless}$
with $N$ particles have been analyzed in Ref.\cite{Rigol}. Following
the approach therein, we let $P$ denote the lowest $N$
eigenfunctions of the Hamiltonian $H_{spinless}$ which can be
obtained by diagonalizing $H_{spinless}$ :
\begin{equation}
P=\left(
\begin{array}{cccc}
P_{11} & P_{12} & \cdots & P_{1N} \\
P_{21} & P_{22} & \cdots & P_{2N} \\
\vdots & \vdots &  & \vdots \\
P_{L1} & P_{L2} & \cdots & P_{LN}%
\end{array}%
\right)_£¬ \label{Pmatrix}
\end{equation}%
where $P_{in}$ are the coefficients of $n$-th single particle state
$|\psi_{n}\rangle=\sum_{i=1}^{L} P_{in} c_i^{\dagger}|0\rangle$.
Then the ground state of the spinless fermion gas is the state with
the lowest $N$ eigenstates of $H_{spinless}$ fully filled, and the
form is:
\begin{eqnarray}
\label{eqn6}
|\Psi^G_{spinless}\rangle&=&\prod^N_{n=1}\sum^L_{i=1}P_{in}c^\dagger_i|0\rangle\notag\\
&=&\sum^{C^N_L}_{s=1}\mathrm{det}(P_s)c^\dagger_s|0\rangle
\end{eqnarray}
where $s$ index the combination formed by taking $N$ numbers from
the set $\Lambda=\{1,\cdot\cdot\cdot,L\}$, $P_s$ is a square matrix
with $N$ ranks that the $N$ rows are taken from $P$ according to the
combination $s$. $c^\dagger_s$ represents
$c^\dagger_{s_1}c^\dagger_{s_2}\cdot\cdot\cdot c^\dagger_{s_N}$, and
$s_i$ is the $i$-th number in the combination $s$. We had assumed
that the numbers in combination are all sorted ascending.
 Now, we can use the reverse mapping of Eq.(\ref{mapping}) to the get
the ground state of the Hamiltonian $H_1$ in the subspace
$M(\vec{\sigma})$
\begin{equation}
\label{eqn3}
|\Psi_{H_1}^G(\vec{\sigma})\rangle=\sum^{C^N_L}_{s=1}\mathrm{det}(P_s)(-1)^{T_q}c^\dagger_{\downarrow
s_q}c^\dagger_{\uparrow s_{\overline{q}}}|0\rangle.
\end{equation}
where $q$ index the combination formed by taking $N_b$ numbers from
the set $\Upsilon=\{1,\cdot\cdot\cdot,N\}$ which means that the
$q_i(i=1,\cdot\cdot\cdot,N_b)$-th site in $\vec{\sigma}$ is occupied
by $\downarrow$-kind fermion and $q_i$ is the $i$-th number in the
combination $q$, $q$ is just another way to index $\vec{\sigma}$.
$c^\dagger_{\downarrow s_q}$ represents $c^\dagger_{\downarrow
s_{q_1}}\cdot\cdot\cdot c^\dagger_{s_{\downarrow q_{N_b}}}$,
$\overline{q}$ represents the combination $\Upsilon-q$, and $T_q$
notes the times of the permutation to put the set
$\{s_{q_1},\cdot\cdot\cdot,s_{q_{N_b}},s_{\overline{q}_1},\cdot\cdot\cdot,s_{\overline{q}_{N_f}}\}$
into $s$. Reminding that the forms of Hamiltonian for $\uparrow$ and
$\downarrow$-kind fermion are the same, the ground states
$|\Psi_{H_1}^G(\vec{\sigma})\rangle$ are $C^{N_b}_N$ degree
degenerate because of $C^{N_b}_N$ different $\vec{\sigma}$.

Supposing the ground state of $H_1$ given by
$|\Phi^{FF}_G\rangle$, the one-particle density matrix function of
boson of the system can be written in the form:
\begin{eqnarray}
\rho_{ij}^{B} &=&\langle\Phi _{G}^{BF}|b^{\dagger}_{i}b_{j}|\Phi
_{G}^{BF}\rangle\notag\\%
&=&\langle\Phi _{G}^{FF}|\prod^{i-1}_{\beta
=1}e^{-i\pi(c_{\beta\downarrow }^{\dagger}c_{\beta\downarrow
}+c_{\beta\uparrow}^{\dagger}c_{\beta\uparrow
})}c^{\dagger}_{i\downarrow}\notag\\
&&\times c_{j\downarrow}\prod^{j-1}_{\gamma
=1}e^{+i\pi(c_{\gamma\downarrow}^{\dagger}c_{\gamma\downarrow}+c_{\gamma\uparrow
}^{\dagger}c_{\gamma\uparrow})}|\Phi _{G}^{FF}\rangle\notag \\
&=&\langle\Phi ^{A}|\Phi ^{B}\rangle ,
\end{eqnarray}%
where $|\Phi _{G}^{BF}\rangle$ is the ground state wave function of
Bose-Fermi mixture and
\begin{equation}
\langle\Phi ^{A}|=\left( c_{i\downarrow}\prod^{i-1}_{\beta =1}%
e^{i\pi (c_{\beta\downarrow }^{\dagger}c_{\beta\downarrow
}+c_{\beta\uparrow }^{\dagger}c_{\beta\uparrow })}|\Phi
_{G}^{FF}\rangle\right) ^{\dagger },\notag
\end{equation}
\begin{equation}
|\Phi ^{B}\rangle=c_{j\downarrow}\prod^{j-1}_{\beta =1}e^{i\pi
(c_{\beta\downarrow }^{\dagger}c_{\beta\downarrow }+c_{\beta\uparrow
}^{\dagger}c_{\beta\uparrow })}|\Phi _{G}^{FF}\rangle.
\end{equation}
In order to calculate $\langle\Phi ^{A}|$ (and $|\Phi ^{B}\rangle$),
it is convenient to use the following identities: \cite{Rigol1}
\begin{equation}
\prod^{i-1}_{\beta =1}e^{i\pi c_{\beta\sigma }^{\dagger}c_{\beta\sigma }}=%
\prod^{i-1}_{\beta =1}[1-2c_{\beta\sigma }^{\dagger}c_{\beta\sigma
}],
\end{equation}%
and
\begin{equation}
\prod^{i-1}_{\beta =1}e^{i\pi c_{\beta\sigma
}^{\dagger}c_{\beta\sigma }}c_{j\sigma }^{\dagger}=(-1)^zc_{j\sigma
}^{\dagger}\prod^{i-1}_{\beta =1}e^{i\pi c_{\beta\sigma
}^{\dagger}c_{\beta\sigma }} ,
\end{equation}
where $z=1$ if $j<i$, otherwise $z=0$. Following the same way shown
above, we can get the one particle density matrix function of
fermion ($\rho^F_{ij}=\langle\Phi
_{G}^{BF}|f^{\dagger}_{i}f_{j}|\Phi _{G}^{BF}\rangle$) and other
quantities such as correlation functions.

\begin{figure}[tbp]
\includegraphics[scale=0.52]{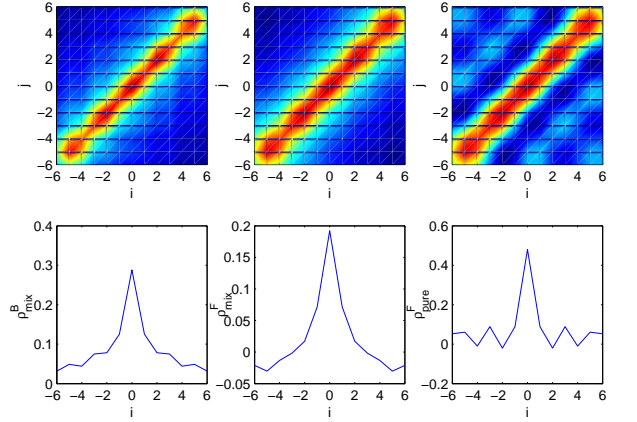}
\caption{(Color online) Top panels: contour plots of the one
particle density matrices. $\rho_{mix}^B$(Left panel),
$\rho_{mix}^F$(middle panel) are the one particle density matrices
for the Bose Fermi mixture with $N_b=3$ and $N_f=2$. Right panel,
the one particle density matrix($\rho_{pure}^F$) for the pure Fermi
gas with $N_f=5$. Bottom panels: corresponding off-diagonal sections
along the anti-diagonal. All the systems are with 13 sites and
$Va^2=0.02t$, where $V$ is the strength of the harmonic trap and $a$
is the lattice spacing.} \label{fig1}
\end{figure}

\section{Hard-core boson-fermion mixture in the harmonic confine potential}
Since the ground state in the hard-core limit has a degeneracy of
$C^{N_b}_N$, for convenience we first consider the case that
$|\Phi^{FF}_G\rangle$ is formed by the summation of
$|\Psi^G_{H_1}(q)\rangle$ with all the degenerate states having the
same weight, {\it i.e.},
\begin{eqnarray}
\label{eqn9}
|\Phi^{FF}_G\rangle&=&\frac{1}{\sqrt{C^{N_b}_N}}\sum_q|\Psi^G_{H_1}(q)\rangle\notag\\
&=&\frac{1}{\sqrt{C^{N_b}_N}}\sum^{C^{N_b}_N}_{q=1}\sum^{C^N_L}_{s=1}\mathrm{det}(P_s)(-1)^{T_q}c^\dagger_{\downarrow
s_q}c^\dagger_{\uparrow s_{\overline{q}}}|0\rangle.
\end{eqnarray}
The ground state $|\Phi^{BF}_G\rangle$ is related to
$|\Phi^{FF}_G\rangle$ by the generalized JWT.  We note that the
above construction is essentially equivalent to the construction of
generalized Bose-Fermi mapping by Girardeau {\it et. al}
\cite{Girardeau07}. Then following the method proposed in the above
section we can work out the density matrix function and show them in
Fig.\ref{fig1}, which are found to fulfil the relation
\begin{equation}
\label{eqn2} \rho^B_{mix}(i,j)=\frac{N_b}{N}\rho^{TG}(i,j)
\end{equation}
according to the data, where $\rho^B_{mix}$ is the one particle
density matrix of boson for the Bose Fermi mixture and
$\rho^{TG}(i,j)$ is the one particle density matrix of pure TG gas
of $N$ bosons obtained by the method proposed by Rigol and Muramatsu
\cite{Rigol}. It is easy to see that the one particle density matrix
of boson decays as the distance grows, while the fermionic one
decays as the distance grows but exhibits typical sign changes due
to the Fermi-Dirac statistics. Same result of Eq.(\ref{eqn2}) for
the continuum systems have been found by Girardeau et. al
\cite{Girardeau07}.

\begin{figure}[tbp]
\includegraphics[scale=0.8]{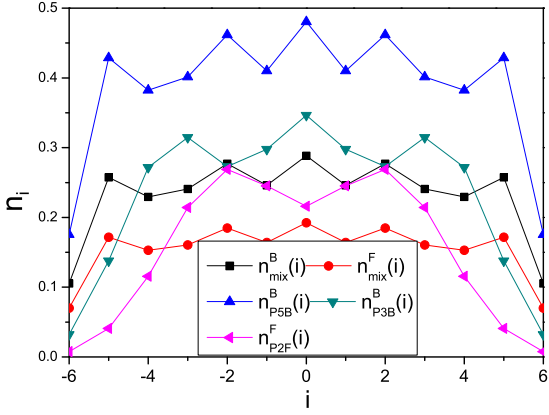}
\caption{(Color online)The density profiles for several systems.
$n_{mix}^{B(F)}(i)$ is bosonic(fermionic) density profile for a Bose
Fermi mixture with $N_b=3$ and $N_f=2$. P$x$B stand for the pure TG
Bose gas with $N_b=x$, and P$x$F stand for the pure free Fermi gas
with $N_f=x$. Again all the systems are with 13 sites and
$Va^2=0.02t$.} \label{fig2}
\end{figure}

The bosonic and fermionic one particle density distributions
$n_{mix}^{B}(i)=\rho _{mix}^{B}(i,i)$ and $n_{mix}^{F}(i)=\rho
_{mix}^{F}(i,i)$ are both proportional to the density
$n^{TG}(i)=\rho^{TG}(i,i)$ of a TG gas of $N$ bosons
\cite{Girardeau}, {\it i.e.,}
\begin{equation}
\label{eqn4}
\frac{n_{mix}^{F}(i)}{N_{f}}=\frac{n_{mix}^{B}(i)}{N_{b}}=\frac{
n^{TG}(i)}{N}=\frac{n^{F}(i)}{N},
\end{equation}
where $n^{F}(i)$ is the density of the noninteracting gas of $N$
fermions in trap which is same to $n^{TG}(i)$\cite{Rigol}. Actually
under the state $|\Phi^{FF}_G\rangle$, the rate of the probabilities
that the site $i$ occupied by the $\uparrow$-kind and
$\downarrow$-kind fermions is $N_b/N_f$. This result means that
there is no phase separation between bosons and fermions. We show
the numerical results of the density profiles in Fig.\ref{fig2}
which agree with theoretic results obtained by Girardeau {\it et.
al} \cite{Girardeau07} and Fang {\it et. al} \cite{Fang}. Comparing
distributions of the TG gas of 3 bosons and the free fermion gas of
2 fermions with the ones for the mixture($n^{B(F)}_{mix}$), we can
see that as the other kind particles adding in, the origin particles
have to hold the higher energy states and the density distributions
become boarder with lower weight.

\begin{figure}[tbp]
\includegraphics[scale=0.8]{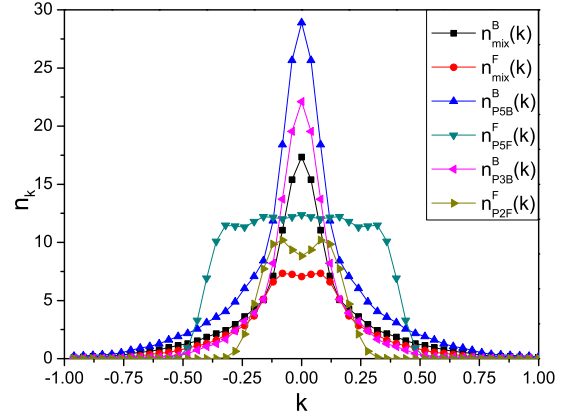}
\caption{(Color online)The momentum distributions for several
systems. The systems are defined by the same way described in
Fig.\ref{fig2}. Notice that the momentum $k$ is in units of $k_L$
which is the wave vector of the optical lattice.} \label{fig3}
\end{figure}

\begin{figure}[tbp]
\includegraphics[scale=0.8]{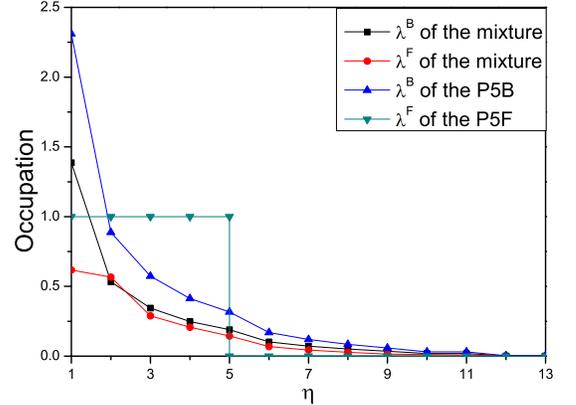}
\caption{(Color online)The occupation of the natural orbitals for
several systems defined by the same way described in
Fig.\ref{fig2}.} \label{fig4}
\end{figure}

The momentum distributions are defined by the Fourier transforms
with respect to $i-j$ of the one particle density matrices with
the form of
\begin{equation}
n^{B(F)}(k)=|\Psi (k)|^{2}\sum^{+\infty }_{n,m=-\infty}%
e^{-ik(n-m)}\rho^{B(F)}(n,m),
\end{equation}
where $\Psi (k)$ is the Fourier transform of the Wannier function,
and $k$ denotes momentum. Since the bosonic one particle density
matrix for the mixture is proportional to the TG one, the bosonic
momentum distribution  for the mixture is also proportional to the
TG one. The numerical results of the momentum distributions are
shown in Fig.\ref{fig3}. The peak structure in the momentum
distribution of boson reflects the bosonic nature of the particle,
and is in contrast with the structure of the momentum distribution
for the equivalent noninteracting fermions. Again the
distributions for the mixture are boarder and lower than the pure
ones because of holding higher energy states.


The natural orbitals ($\phi _{\eta }^{B(F)}(i)$) are defined as the
eigenfunctions of the one particle density matrix\cite{Penrose}:%
\begin{equation}
\sum^{M}_{j=1}\rho ^{B(F)}(i,j)\phi _{\eta }^{B(F)}(j)=\lambda
_{\eta }^{B(F)}\phi _{\eta }^{B(F)}(i)
\end{equation}%
and it can be understood as being effective one particle states with
occupation $\lambda _{\eta }$. In Fig.\ref{fig4} we show the
occupations for boson and fermion. The occupations are plotted as a
function of the orbital numbers $\eta$, and they are ordered by
starting from the highest one. As the one particle density matrix of
boson of the mixture is proportional to the TG one, the natural
orbitals of boson of the mixture are the same as the TG one with the
occupations fulfilling the following relations
\begin{equation}
\lambda^B_\eta=\frac{N_b}{N}\lambda^{TG}_\eta,
\end{equation}
where $\lambda^{TG}_\eta$ denotes the occupation of $\eta$-th
natural orbital for a pure TG gas composed of $N$ hard-core bosons.
The peak on the lowest orbital of boson is the feature of the boson.
As for the fermion, the occupation is no longer the step function
with lowest $N$ orbitals fully filled as the distribution of the
pure noninteracting $N$ fermions. And there is no peak at the lowest
orbital.

\section{The property of the system with large but finite BB and BF repulsion}

As we have discussed in section II, the ground state in the
hard-core limit has a huge degeneracy. However, we expect that the
degenerate ground state would be lifted when the on-site
interactions deviate the infinite limit. Next we consider the case
with $U_{BB}$ and $U_{BF}$ being large but finite, for which the
second quantized Hamiltonian of $H$ is the standard Hubbard model of
Bose-Fermi mixture with the form of Eq.(\ref{BFHubbard}). In the
situation that $U_{bb}$ and $U_{bf}$ are still large, the state with
double occupancy on the same site is a high-energy state and we can
use the standard projection method to derive the low-energy
effective Hamiltonian of the system which is given by
\begin{eqnarray}
\label{eqn7}
H_s&=&-t\sum^{L-1}_{i=1}(b^\dagger_ib_{i+1}+f^\dagger_if_{i+1}+H.c.)\notag\\
&&+Va^{2}\sum^{L}_{i=1}i^{2}n^b_{i}+Va^{2}\sum^{L}_{i=1}i^{2}n^f_{i}\notag\\
&&-\frac{4t^2}{U_{bf}}\sum^{L-1}_{i=1}[(S^x_i S^x_{i+1}+S^y_i
S^y_{i+1})-S^z_i
S^z_{i+1}]\notag\\
&&-\frac{2t^2}{U_{bb}}\sum^{L-1}_{i=1}n_i^bn^b_{i+1} \label{Htj}
\end{eqnarray}
with $S^\dagger_i=b^\dagger_if_i, S_z=(n_i^b-n_i^f)/2$.
In the limit $U_{bb},U_{bf}\rightarrow\infty$, the last two
summation terms in the Hamiltonian vanish and the system reduces
back to Eq.(\ref{eqn4}) which we studied in the last section. As
$U_{bb}$ and $U_{bf}$ become finite but large, the last two
summation terms can be viewed as perturbations to the system, so one
can expect that they wouldn't cause significant changes to some
properties of the total system such as ground state energy and total
density profile (see Fig.\ref{fig6}b) because of terms of
${t^2}/{U_{bf}}$ and ${t^2}/{U_{bb}}$ being very small for large
$U_{bb}$ and $U_{bf}$. Nevertheless, a significant effect induced by
these small terms is the lift of the degeneracy of the ground states
and the true ground state would be a recombination of $C_N^{N_b}$
degenerate states with the weights of states to be determined by
minimization of the energy due to perturbation terms.

One observes that terms of $t^2/U_{bf}$ lead to an effective
isotropic antiferromagnetic exchange interactions between ``spins"
(bosons or fermions) on neighboring sites to lower the ground state
energy. On the other hand, terms of $t^2/U_{bb}$ produce an
effective attractive interactions between neighboring bosons and
thus states with all the bosons concentrated together have lower
energy. Consequently, the relative distribution for the bosons or
fermions will be changed and determined by terms of $t^2/U_{bb}$ and
$t^2/U_{bf}$ to further lower the ground state energy. However, one
can expect that the total distribution shall not be changed too much
because the terms of $t^2/U_{bb}$ and $t^2/U_{bf}$ are very small in
comparison with hopping terms.  As the system Eq.(\ref{Htj}) has no
analytical results any more when the interaction parameters
$U_{bb},U_{bf}$ being finite, as an approximation, we can treat the
charge part and spin part separately. Then the charge part, which
does not distinguish bosons or fermions, is determined by the matrix
$P$ according to Eq.(\ref{Pmatrix}) with the state given by
Eq.(\ref{eqn6}), whereas the spin part, which decides the weight of
states with different spin configurations, is determined by
Hamiltonian with only $U_{bf}$ and $U_{bb}$ terms in
Eq.(\ref{eqn7}). Considering that, with the presence of the harmonic
trap, the particles mainly concentrate at the center of the trap
with holes around, the spin part is defined on the system with $N_b$
bosons and $N_f$ fermions on $N_b+N_f$ sites. This approximation is
similar to the spin charge separation approximation.
\begin{figure}[tbp]
\includegraphics[scale=0.8]{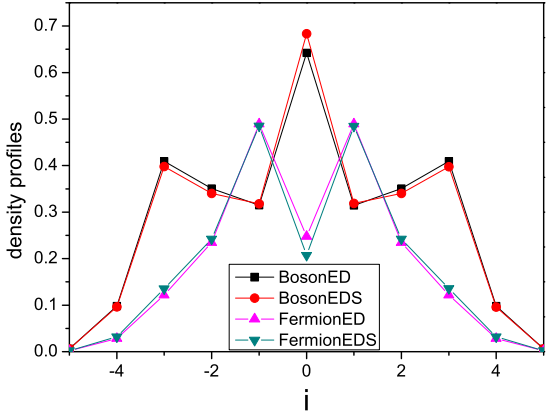}
\caption{(Color online)The density profiles of the Bose Fermi
mixture with 3 boson and 2 fermion on 11 sites, the trap strength
$Va^2=0.2t$, $U_{bb}=2000t$ and $U_{bf}=200t$. 'ED' stands for exact
diagonalization. 'EDS' stands of the results gained by spin charge
separation approximation with the spin part (weights of subspace
ground states) is determined by exact diagonalization.} \label{fig5}
\end{figure}

First, we consider the situation $U_{bb}\gg U_{bf}\gg1$. In this
limit, the terms of $U_{bf}$ dominate and we can set
$U_{bb}=\infty$. Because of the spin fluctuation terms of
$S_i^xS_j^x + S_i^yS_j^y$, the off-diagonal terms appear between
different subspace $M(q)$, and the ground state of the system
becomes complicated. Actually in the limit $U_{bb}=\infty$, the
Hamiltonian $H_{Hub}$ can be mapped to the Fermi Hubbard model by a
JWT similar to Eq.(\ref{JWT}) and they have the same thermodynamic
properties \cite{Chen09}. As for the ground state properties, it can
be worked out by the mapping from the fermi Hubbard model. When
$U_{bb}$ is away from the infinite limit, there are no analytical
results. In Fig.\ref{fig5} we show the density profiles in the
situation $U_{bb}\gg U_{bf}\gg1$ with spin charge separation
approximation. From the data, the total distribution is the same
with that of the pure TG boson gas. Because of the effective
antiferromagnetic exchanges between bosons and fermions, the
specie-dependent distributions for bosons and fermions exhibit quite
different behavior with alternating peaks. To check the validity of
this approximation method, in Fig.\ref{fig5} we also show some
results of small system of Bose-Fermi Hubbard model gained by exact
diagonalization method \cite{Roth}.  We can see that the results
gained by the approximation agree well with the results obtained by
exact diagonalization and the density distributions of the total
particle are almost the same for the data obtained from both methods
(the difference $<10^{-2}$).

\begin{figure}[tbp]
\includegraphics[width=7cm, height=6cm, bb=25 20 303 235]{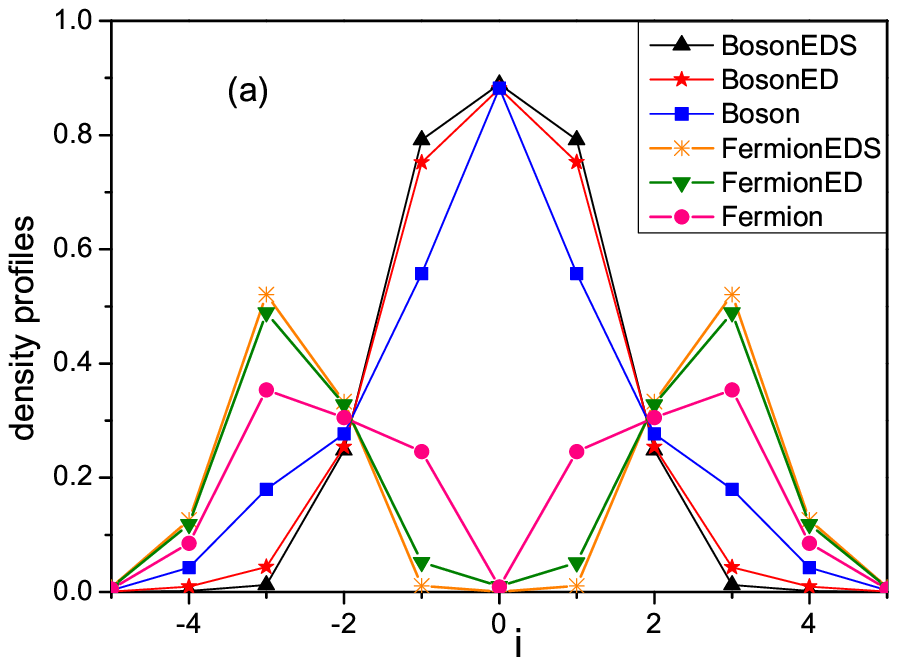}
\includegraphics[width=7cm, height=5cm, bb=25 20 303 235]{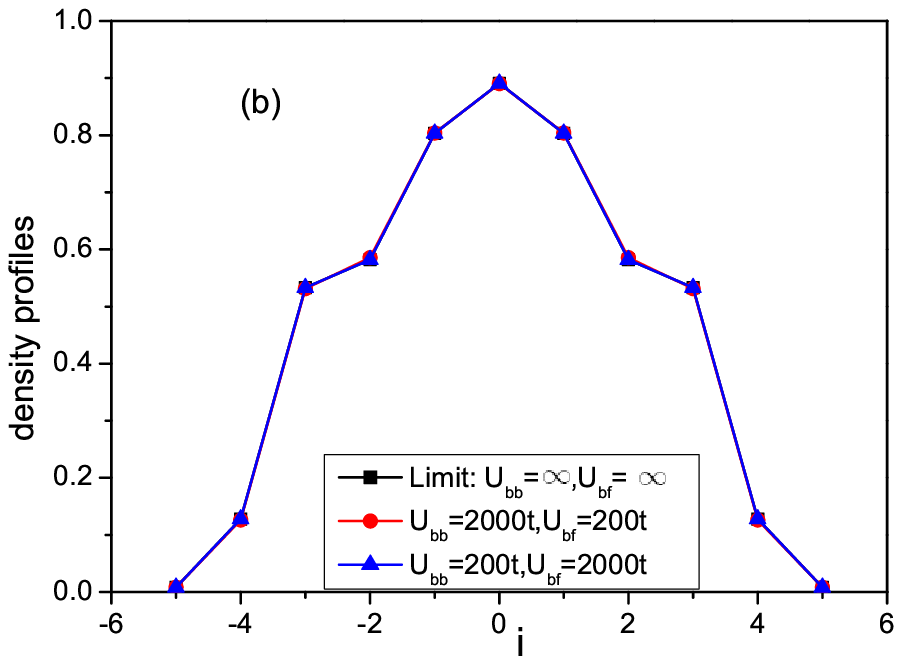}
\caption{(Color online)(a):The density profiles of the Bose Fermi
mixture with 3 boson and 2 fermion on 11 sites, the trap strength
$Va^2=0.2t$, $U_{bb}=200t$ and $U_{bf}=2000t$. The curve with symbol
square (circle) is the density profile of the boson (fermion) for the
Bose Fermi mixture calculated from the ground state
$|\Phi^{FF}_G\rangle$ in Eq.\ref{eqn5}. (b): The total density
profiles of different $U_{bb},U_{bf}$ for the Bose Fermi mixture
with 3 boson and 2 fermion on 11 sites, the trap strength
$Va^2=0.2t$. The curve with symbol square is the total density
profile calculated from the ground state $|\Phi^{FF}_G\rangle$ in
Eq.\ref{eqn9}.} \label{fig6}
\end{figure}

Next we consider the situation $U_{bf}\gg U_{bb}\gg1$. Then the
$U_{bb}$ terms dominate and for simplicity we can first set
$U_{bf}=\infty$. Although the $U_{bb}$
term is diagonal in the Hilbert space $\cup_qM(q)$,
the states $|\Psi^G_{H_1}(q)\rangle $ (Eq.(\ref{eqn3})) with different spin configurations
are not degenerate any more for
different $q$ because of the $U_{bb}$ term. Since the $U_{bb}$ terms
tend to make the boson concentrated together, the ground states of the
configuration $q$ with all the boson staying together have the
lowest energy. Then the ground states have the degeneracy of $C^1_{N_f+1}=N_f+1$.
We suppose that the ground state
$|\Phi^{FF}_G\rangle$ is formed by these degenerate states in
subspace $M(q)$ with all the degenerate states having the same weight,
say:
\begin{eqnarray}
\label{eqn5}
|\Phi^{FF}_G\rangle&=&\frac{1}{\sqrt{N_f+1}}\sum_{q'=1}^{N_f+1}|\Psi^G_{H_1}(q')\rangle\\
&=&\frac{1}{\sqrt{N_f+1}}\sum^{N_f+1}_{q'=1}\sum^{C^N_L}_{s=1}\mathrm{det}(P_s)(-1)^{T_{q'}}c^\dagger_{\downarrow
s_{q'}}c^\dagger_{\uparrow s_{\overline{q'}}}|0\rangle.\notag
\end{eqnarray}
Then we can get the density matrices and other quantities. In
Fig.\ref{fig6}a, we show the density of the system form the state
$|\Phi^{FF}_G\rangle$. We can see that there is a phase separation
in the system with bosons are in the middle of the trap and fermions
surround them. As $U_{bf}$ is away from the limit $\infty$, the $N_f+1$
fold degeneracy of the ground state is split.
For comparison,t he density profiles in the limit $U_{bf}\gg U_{bb}\gg1$ with spin
charge separation approximation and exact diagonalization are also shown
in Fig.\ref{fig6}a. It is clear that
the results obtained by spin charge separation approximation agree well
with the ED ones, however the results gained by the state
$|\Phi^{FF}_G\rangle$(Eq.(\ref{eqn5})) do not agree very well with
the others because the $U_{bf}$ terms are neglected. But the results
gained by the state $|\Phi^{FF}_G\rangle$(Eq.(\ref{eqn5})) and spin
charge separation approximation both indicate that there is a phase
separation in the system with bosons located in the middle of the trap
and fermions surrounded. Again the density distributions of the
total particle are almost the same from the data for the three
methods (the difference $<10^{-2}$).

\section{Summary}
In conclusions, we have studied in detail the ground state
properties of the mixture of the hard-core bosons and noninteracting
fermions with point hard-core boson-boson and boson-fermion
interactions in the 1D optical lattice with harmonic confine
potential. Using extended Jordan-Wigner transformations, we
calculate the density matrix, then we yield the density profiles,
momentum distribution, the natural orbitals and its occupations. We
also discuss the property of the system with large but finite
interactions. We find that, despite the total density distribution
not sensitive to relative strengths of $U_{bf}$ and $U_{bb}$, the
boson and fermion distributions rely on $U_{bb}\gg U_{bf}\gg1$ or
$U_{bf}\gg U_{bb}\gg1$. We hope that our study could be helpful for
the experimental achievement of the ultracold boson-fermion mixtures
with hard-core BB, BF interactions in optical lattices.

\begin{acknowledgments}
This work was supported by NSF of China under Grants No. 10821403
and No. 10974234, programs of Chinese Academy of Sciences, 973 grant
No. 2010CB922904 and National Program for Basic Research of MOST.
\end{acknowledgments}


\begin{references}
\bibitem{note1} The on-site constraints can be realized with generalized constrained
operators
$\bar{b}_{i}^{\dagger}=b_{i}^{\dagger}(1-f_{i}^{\dagger}f_{i})$
 and
 $\bar{f}_{i}^{\dagger}=f_{i}^{\dagger}(1-b_{i}^{\dagger}b_{i})$,
 where $b_{i}^{\dagger}$ is the hard-core boson operator which can
 be constructed by the conventional Jordan-Wigner transformation.
 For convenience, we still use $b_{i}^{\dagger}$ and $f_{i}^{\dagger}$ instead of
 $\bar{b}_{i}^{\dagger}$ and $\bar{f}_{i}^{\dagger}$ to represent generalized constrained
operators in the following context whenever we discuss the model in
the hard-core limit.
\end{references}
\end{document}